\documentclass[a4paper,fleqn]{cas-dc}

\usepackage[authoryear]{natbib}

\usepackage{amssymb}
\usepackage{amsmath}
\usepackage[switch]{lineno}
\usepackage{hyperref}
\usepackage{xurl}
\usepackage[normalem]{ulem}

\usepackage[dvipsnames]{xcolor}
\usepackage{verbatim}

\usepackage{graphicx}
\usepackage{subcaption}

\providecommand{\apj}{ApJ}                
\providecommand{\apjl}{ApJ}               
\providecommand{\apjs}{ApJS}              
\providecommand{\aap}{A\&A}               
\providecommand{\jcp}{J.~Chem.~Phys.}     
\providecommand{\jqsrt}{J.~Quant.~Spectrosc.~Radiat.~Transf.} 
\providecommand{\mnras}{MNRAS}            
\providecommand{\nat}{Nature}             

\definecolor{amethyst}{rgb}{0.6, 0.4, 0.8}

\definecolor{amber}{rgb}{1.0, 0.75, 0.0}

\definecolor{awesome}{rgb}{1.0, 0.13, 0.32}

\ExplSyntaxOn
\cs_gset:Npn \__first_footerline:
  {
    \group_begin:
    \small
    \sffamily
    \__short_authors:
    \group_end:
  }
\ExplSyntaxOff

\begin{document}
\let\WriteBookmarks\relax
\def\floatpagepagefraction{1}
\def\textpagefraction{.001}

\shorttitle{The \texttt{CosmicPAH\textendash IRDB} web application}    

\shortauthors{de~Bentzmann \textit{et al.}}  

\title [mode = title]{\texttt{CosmicPAH-IRDB}: A Web Application for Including Anharmonicity in Simulated PAH Spectra for Astrophysics}




%

\author[irap]{Louan de~Bentzmann}[orcid=0009-0005-4421-948X]
\credit{Software, Methodology, Formal analysis, Data curation, Visualization, Writing – original draft }
\author[irap]{Christine Joblin}[orcid=0000-0003-1561-6118]
\credit{Conceptualization, Supervision, Validation, Resources, Project administration, Funding acquisition, Writing – original draft, Writing – review and editing}
\cormark[1]
\ead{christine.joblin@cnrs.fr} 
\author[irap]{Karine Demyk}[orcid=0000-0002-5019-8700]
\credit{Investigation, Formal analysis, Methodology, Data curation, Resources, Writing – review and editing}
\author[inaf-oaca,irap]{Giacomo Mulas}[orcid=0000-0003-0602-6669]
\credit{Investigation, Methodology, Supervision, Data curation, Resources, Writing – review and editing}
\author[lcar]{Dominique Toublanc}[orcid=0000-0002-2556-051X]
\credit{Software, Methodology, Formal analysis}
\author[irap]{Vincent Marty}
\credit{Software, Methodology}
\author[irap]{Axel Pérignon}
\credit{Software, Methodology}
\author[irap]{Mickaël Boiziot}[orcid=0000-0003-0298-9820]
\credit{Software}
\author[irap]{Jean-Michel Glorian}[orcid=0000-0002-2467-0506]
\credit{Software}

\affiliation[irap]{organization={Institut de Recherche en Astrophysique et Plan\'etologie (IRAP), Universit\'e de Toulouse, CNRS, CNES},
                addressline={9, Avenue du Colonel Roche},
                city={Toulouse},
                postcode={F-31028},
                country={France}}

\affiliation[inaf-oaca]{organization={Istituto Nazionale di Astrofisica (INAF), Osservatorio Astronomico di Cagliari},
            addressline={Via della Scienza 5}, 
            city={Selargius},
            postcode={09047}, 
            state={CA},
            country={Italy}}
            
\affiliation[lcar]{organization={Laboratoire Collisions Agrégats Réactivité (LCAR/FeRMI), Université de Toulouse, CNRS},
                addressline={118 Route de Narbonne},
                city={Toulouse},
                postcode={F-31062},
                country={France}}

















\begin{abstract}
Advances in infrared observations with the James Webb Space Telescope (JWST) call for new tools to exploit the information associated with the Aromatic Infrared Bands (AIBs), now observed in both distant galaxies and the embryos of planetary systems. The AIBs contain information about their carriers as well as the excitation conditions, making them powerful probes of the chemical and physical conditions of their environments -- if their spectral information can be decoded.

While large datasets of theoretical infrared spectra are becoming available, efforts have primarily focused on the contribution of chemical diversity to the AIBs. However, the results remain limited by the impact of the excitation process on the band profiles. The AIB carriers, thought to be polycyclic aromatic hydrocarbons (PAHs), are hot molecules, which affects their spectral characteristics.

We report here our approach to quantify these effects empirically by collecting IR spectra -- both experimental and theoretical -- and studying the evolution of band positions and widths with temperature. We present our methodology, including the development of the \texttt{CosmicPAH\textendash IRDB} web application and spectral database, along with the associated \texttt{cosmicPAHmfit} multi-component spectral fitting tool. Our philosophy emphasizes open sharing, user-friendly access, and machine readability. The web application, tool and their update are accessible via the \texttt{Cosmic PAH portal} (\url{https://cosmic-pah.irap.omp.eu/}).

\end{abstract}






\begin{NoHyper}
\maketitle
\end{NoHyper}


\section{Introduction}\label{sec-intro}

Observations by the James Webb Space Telescope (JWST) provide an unprecedented view of the infrared features associated with interstellar and circumstellar matter.  Among these features, the Aromatic Infrared Bands (AIBs) -- a set of emission bands with main components at 3.3, 6.2, 7.7, 8.6, 11.3, and 12.7\,$\mu$m -- stand out prominently. Approximately forty years ago, it was proposed that these features arise from the radiative cooling of large carbonaceous molecules, specifically polycyclic aromatic hydrocarbons (PAHs), following the absorption of ultraviolet (UV) photons from stars \citep{leger1989, allamandola1989}. The exceptional quality of JWST observations now demands detailed simulated spectra for comparison with these observations.

The principle of infrared radiative cooling was described in earlier studies supported by laboratory investigations \citep[e.g.,][]{brenner1992}. However, emission spectra for high-excitation energies -- such as those achieved by the absorption of a UV photon -- remain scarce \citep{brenner1992, cook1996}. While cryogenic storage rings offer new possibilities for studying the dynamics of radiative cooling, including IR cooling for PAHs \citep{stockett2019_ircooling, stockett2020_ircooling}, spectroscopic studies following the pioneering work of the 1990s are still awaited.

Simulating the IR emission spectra of PAHs in astrophysical environments for comparison with observed AIB spectra requires several key ingredients, as discussed in earlier studies \citep{cook1998_apj, pech2002}. These studies emphasize the need for experimentally obtained spectral parameters characteristic of highly excited PAHs. Empirical linear laws quantifying the evolution of band positions and widths with temperature were first reported by Joblin et al. \citep{joblin1995} for thermally excited PAHs up to temperatures of $\sim$900~K. These laws have since been used in PAH emission models to compare simulated spectra with observed AIB spectra \citep{verstraete2001,pech2002}.

However, earlier works were limited by the availability of spectral data, covering only a few species and bands. Most modelling studies of the AIBs have since focused on global chemical diversity, including charge state and size. Such analyses are conducted using the \href{https://www.astrochemistry.org/pahdb/}{NASA Ames PAH IR Spectroscopic Database}, 
an extensive database of theoretical spectra, but simplify or overlook the impact of the emission mechanism on spectral features \citep{boersma2014, bauschlicher2018, maragkoudakis2025, ricca2026, maragkoudakis2026}. A number of studies have emphasised the need to include temperature effects on band characteristics in the emission-cascade treatment, which appears crucial not only in the CH stretch range \citep{chen2018ApJS, lemmens2021} but also in the studies of oxygen-functionalized PAHs \citep{mishra2025}. We recently demonstrated the importance of such an approach in the assignment of the carriers of the 3.4~$\mu$m band, a satellite of the main 3.3~$\mu$m aromatic band \citep{demyk2026}.

The evolution of IR spectra with temperature is related to intermode and intramode anharmonicity \citep{cook1998_jpca}. Harmonic spectra, which are readily available in databases (the \href{https://www.astrochemistry.org/pahdb/}{NASA Ames PAH IR Spectroscopic Database} and the \href{https://cosmicpah-qcals.oa-cagliari.inaf.it/}{Theoretical spectral database of polycyclic aromatic hydrocarbons}), are obtained by assuming that vibrational modes are independent. In reality, these modes are coupled even at 0~K. Theoretical calculations have shown that anharmonic effects lead to the appearance of numerous combination bands \citep{mackie2018a,mulas2018, esposito2024}. At higher temperatures, this effect is amplified as more vibrational bands become populated, increasing the number of transitions that contribute to the IR spectrum \citep{mackie2018b, chen2018AandA, chakraborty2021}.

To quantify empirically anharmonic effects in the IR spectra of hot PAHs in a systematic manner, we have developed a spectral fitting tool and a dedicated web application that can be utilized not only by our group but also by the broader community. The objective of this article is to present:

\begin{itemize}
    \item \textbf{\texttt{cosmicPAHmfit}}, a multi-component spectral fitting tool that is key to band analysis in experimental or theoretical spectra of hot PAHs. This tool also enables the derivation of empirical anharmonicity functions and associated factors.
    \item \textbf{\texttt{CosmicPAH\textendash IRDB}}, a web application that collects experimental and theoretical spectra to quantify anharmonicity. The web application also includes empirical anharmonicity factors and provides an overplot functionality, which is convenient for spectral analysis and band assignment.
\end{itemize}

This article will guide the reader through the different steps involved in this work, up to the illustration of the use of spectral data for modelling the emission of UV-excited PAHs. The spectra of both pyrene (C$_{16}$H$_{10}$) and hexahydropyrene (C$_{16}$H$_{16}$, 6H-pyrene) are used throughout the manuscript. We highlight the approximations implemented so far and discuss perspectives for future evolution.

\section{From infrared spectra of PAHs to spectral parameters}

\subsection{Spectroscopic datasets of hot PAHs}\label{sec-spectro}

\begin{figure*}
    {\centering
    \includegraphics[height=5.4 cm]{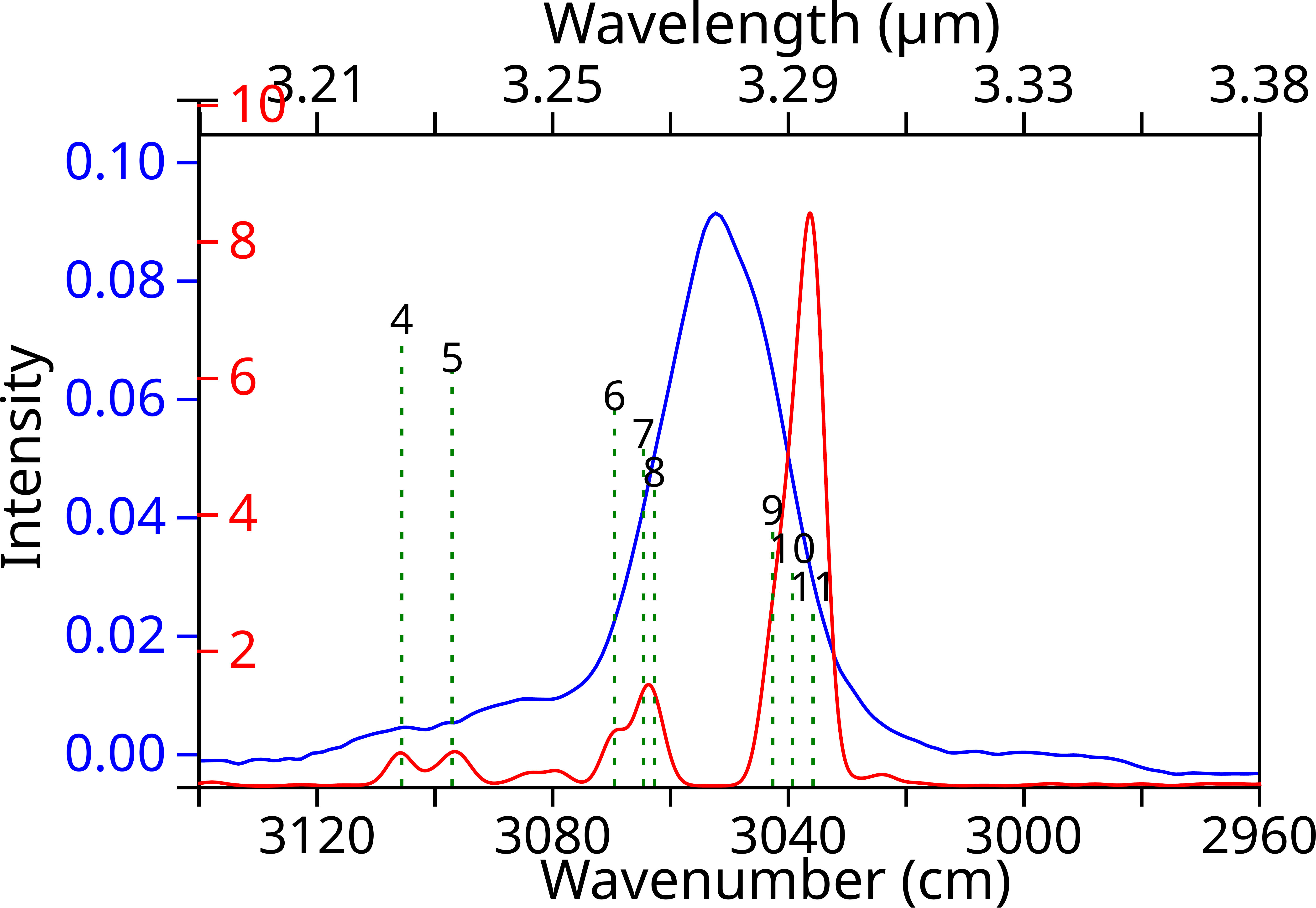}
    \includegraphics[height=5.4 cm]{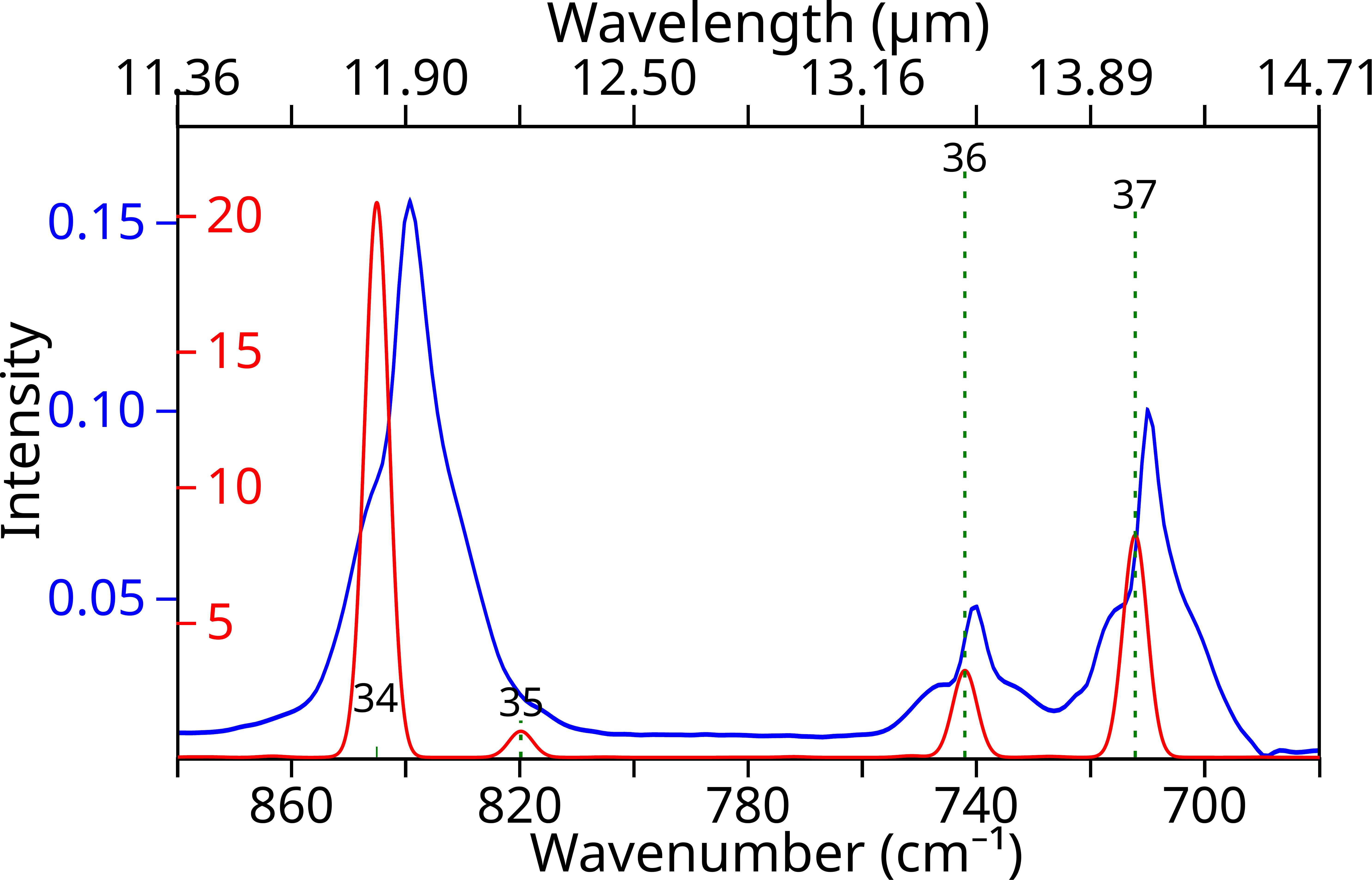} 
    \begin{tabular}{cccl}
         \\
        \hline
        Index & Wavenumber & Intencity & State \\
        &(cm$^{-1}$)& (km/mol)&\\
        \hline
        4 & 3105.37& 1.700 & $\nu_{34}+\nu_{46}+\nu_{49}(0.31),
        \nu_{44}+\nu_{57}(0.30), \nu_{10}+\nu_{52}+\nu_{57}(0.15), $\\
        &&&$\nu_{12}+\nu_{44}+\nu_{57}(0.04), \nu_{34}+\nu_{57}+\nu_{62}(0.03),$\\
        &&&$ \nu_{12}+\nu_{34}+\nu_{46}+\nu_{49}(0.03), \nu_{23}(0.02), \nu_{25}(0.02)$ \\
        5 & 3096.78 & 1.821 & $\nu_{4}+\nu_{45}(0.73), \nu_{4}+\nu_{12}+\nu_{45}(0.09), 2\nu_{32}+\nu_{45}(0.06), \nu_{42}(0.05)$ \\
        6 & 3069.24 & 3.965 & $\nu_{21}+\nu_{56}+\nu_{66}(0.49), \nu_{45}+\nu_{56}(0.23), \nu_{23}(0.09), $\\
        &&&$\nu_{12}+\nu_{21}+\nu_{56}+\nu_{66}(0.05), \nu_{44}+\nu_{57}(0.04)$ \\
        7 & 3064.29 & 4.277 & $\nu_{29}+\nu_{45}+\nu_{53}(0.32), \nu_{21}+\nu_{56}+\nu_{66}(0.28), \nu_{45}+\nu_{56}(0.12), \nu_{23}(0.10), $\\
        &&&$\nu_{12}+\nu_{29}+\nu_{45}+\nu_{53}(0.04), \nu_{44}+\nu_{57}(0.03), \nu_{12}+\nu_{21}+\nu_{56}+\nu_{66}(0.03)$ \\
        8 & 3062.45 & 4.260 & $\nu_{29}+\nu_{45}+\nu_{53}(0.48), \nu_{21}+\nu_{56}+\nu_{66}(0.12), \nu_{23}(0.10), \nu_{45}+\nu_{56}(0.09), $\\
        &&&$\nu_{12}+\nu_{29}+\nu_{45}+\nu_{53}(0.06), \nu_{21}+\nu_{29}+\nu_{53}+\nu_{66}(0.03), \nu_{44}+\nu_{57}(0.03) $\\
        9 & 3042.38 & 9.577 & $\nu_{42}(0.58), \nu_{43}(0.27), \nu_{12}+\nu_{42}(0.05)$ \\
        10 & 3039.01 & 16.663 & $\nu_{23}(0.48), \nu_{45}+\nu_{56}(0.30), \nu_{12}+\nu_{23}(0.05), \nu_{4}+\nu_{51}+\nu_{65}(0.04), $\\&&&$\nu_{12}+\nu_{45}+\nu_{56}(0.04)$\\
        11 & 3035.49 & 37.713 & $\nu_{43}(0.62), \nu_{42}(0.25), \nu_{12}+\nu_{43}(0.06)$\\
        34 & 844.7 & 108.841 & $\nu_{67}(0.91)$\\
        35 & 819.53 & 5.117 & $\nu_{32}(0.91)$ \\
        36 & 741.73 & 16.973 & $\nu_{68}(0.91)$\\
        37 & 711.89 & 43.592 & $\nu_{69}(0.91)$ \\
        \hline \\
    \end{tabular}
    \includegraphics[height=5.4 cm]{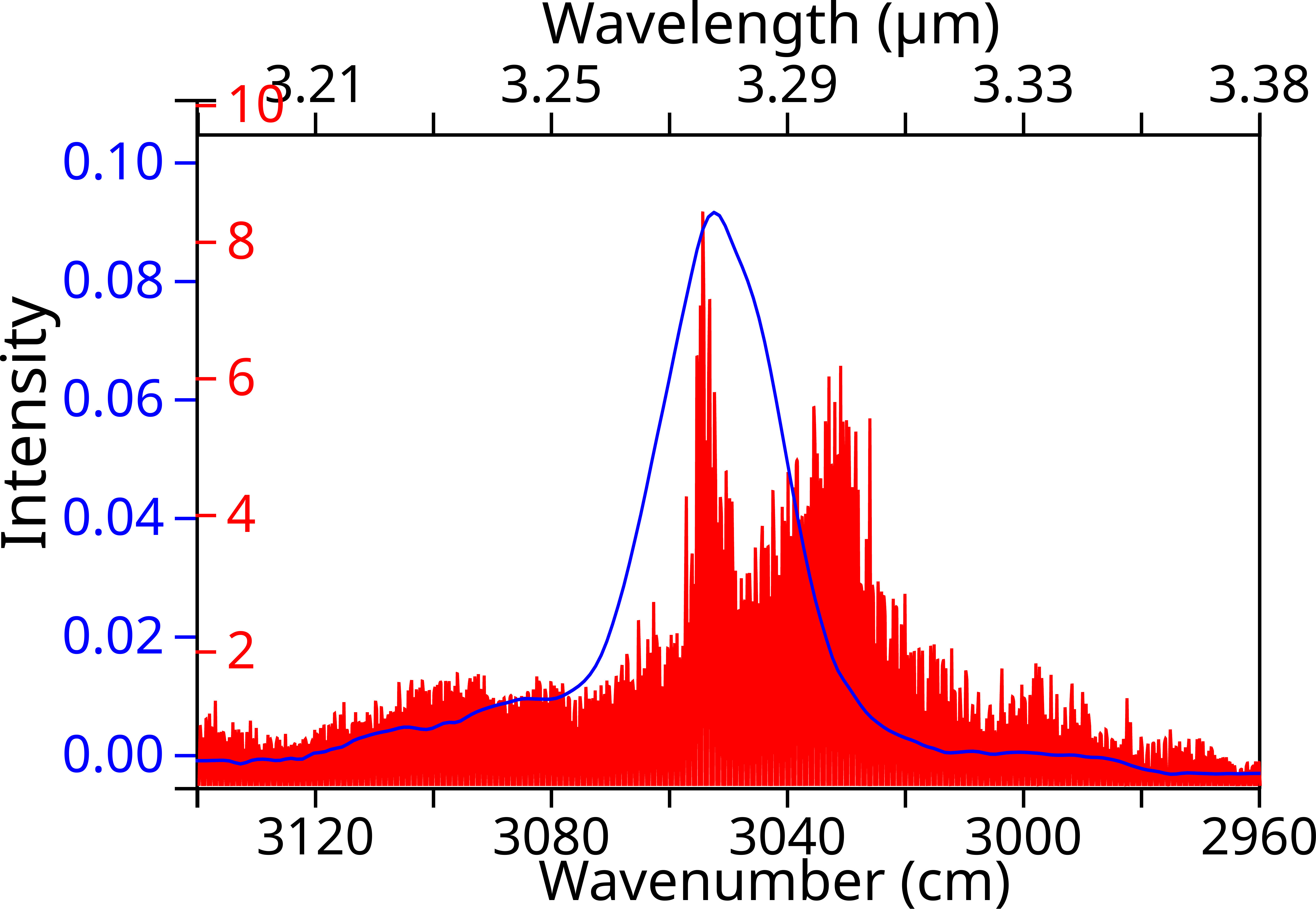}
    \includegraphics[height=5.4 cm]{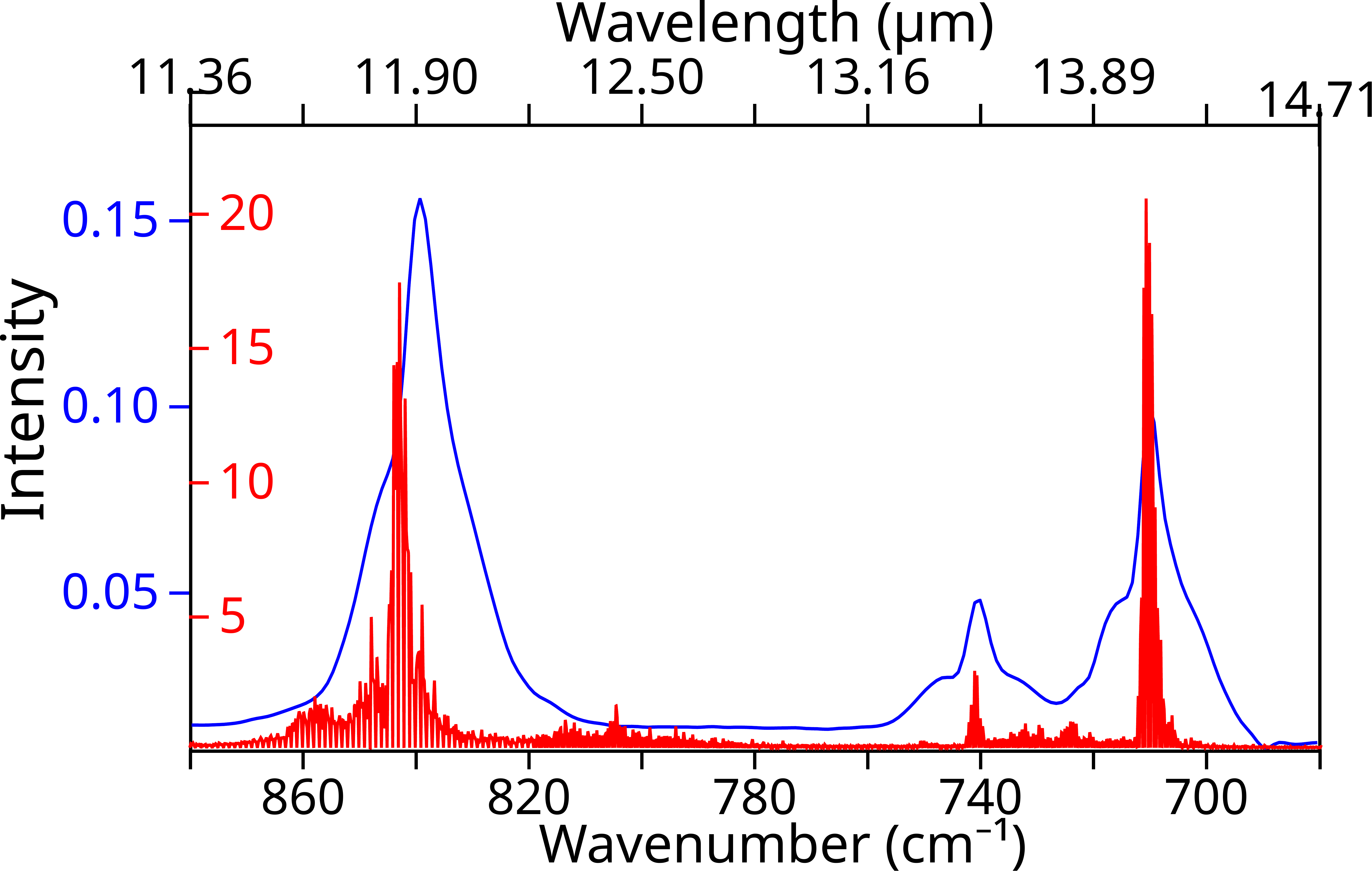}
    \caption{IR spectrum of pyrene, focusing on the strongest CH bands in the 3.3~$\mu$m and 12~$\mu$m ranges. The experimental gas-phase spectrum at 523~K \citep{demyk2026} is compared with AnharmoniCaOs spectra calculated at 0~K \citep[top graphs,][]{mulas2018} and 532~K \citep[bottom graphs,][]{chakraborty2021}. The graphs can be regenerated using the following links: 
   \href{https://cosmicpah-irdb.irap.omp.eu/science/Pyrene/Gas/2/sample/?overplot-tab\#/3/24/5/1.5/3140/720/1500/500/523/}{Upper graphs}  {\href{https://cosmicpah-irdb.irap.omp.eu/science/Pyrene/Gas/2/sample/?overplot-tab\#/3/26/5/1.5/3140/720/1500/500/523/523}{Bottom graphs} 
    \label{fig:comppy} }}}
\end{figure*} 

Our developments were motivated by the acquisition of new datasets on the evolution of PAH IR spectra with temperature using the FTIR setup ESPOIRS. The measurements were performed under thermal equilibrium conditions at well-defined temperatures, either in the gas phase or in the condensed phase \citep{demyk2026}. In the condensed phase, the sample is diluted in a pellet of matrix material (e.g., KBr) that remains transparent and chemically non-reactive over the temperature range of 14~K to 723~K for KBr \citep{chakraborty2019}. In the gas phase, the PAH sample (solid at room temperature) is placed in an oven, and the resulting gas-phase species are thermalised by collisions with a N$_2$ buffer gas at a pressure of approximately 200~mbar at room temperature. Under these conditions, the population of rotational states is in thermal equilibrium at the same sample temperature as vibrational states. This differs from astronomical environments, where PAHs are rotationally much colder. In the condensed phase, molecular rotation is blocked, but band positions and profiles are perturbed relative to the isolated molecule due to interactions with the KBr matrix and nearby PAH molecules.

In addition to experimental data, anharmonic theoretical spectra can be calculated at various temperatures. These calculations are computationally much more expensive than the commonly available harmonic spectra. Anharmonic spectra can be obtained using frameworks such as GVPT2, with tools like \href{https://sourceforge.net/projects/anharmonica/}{AnharmoniCaOs} \citep{mulas2018} or \textsc{SPECTRO} \citep{mackie2015}, both at 0~K and -- with a much larger computational cost -- at medium temperatures \citep[up to $\sim$550~K using AnharmoniCaOs;][]{chakraborty2021}. At 0~K anharmonic transitions, while already significantly more numerous than in the harmonic approximation, can still be easily enumerated and listed. At significantly higher temperatures their number gets so large that, while being still possible in principle, it becomes impractical, making the calculation slows down just to write them in outputs of hundreds of gigabytes. In these cases, one just saves discretised spectra, with the cumulative intensity in bins of a chosen width, corresponding to the resolution of the calculation. It is then left to the user to further convolve these binned spectra with a custom profile (e.g. a Gaussian with a fixed width) to ease visualisation.

Anharmonic spectra can also be generated via Born-Oppenheimer Molecular Dynamics (BOMD) simulations, using different backends such as Density Functional Theory (DFT) or DFT-based Tight-Binding (DFTB) to compute forces and electric properties on the fly. These calculated spectra also contain information on the band width and can be directly compared to experimental spectra. These methods offer varying computational costs and accuracies. Notably,
the computational cost of BOMD simulations does not increase significantly with the target temperature, so that, when feasible for one temperature for a given species, they can, in principle, be performed at any other desired one \citep{joalland2010,simon2011, VanOanh2012, chen2019, chakraborty2021}. 

Figure~\ref{fig:comppy} shows the comparison of the main IR bands of pyrene (i.e., C-H stretches, C-H out-of-plane bends) from experimental gas-phase spectra and anharmonic calculations. The detailed assignment provided by the anharmonic calculations at 0~K illustrates that the C-H out-of-plane bending vibrations involve only fundamental transitions at 0~K. In contrast, a diversity of combination bands, in addition to the fundamental bands, is present at higher frequencies in the region of the C-H stretching vibrations (see list of transitions in Figure~\ref{fig:comppy}). 
As previously mentioned, at higher temperatures it is impractical to list all individual hot bands as was done at 0~K (upper panels of Figure~\ref{fig:comppy}), so just a discretised histogram spectrum is shown, with bins of finite, fixed size (bottom panel of Figure~\ref{fig:comppy}).

\subsection{Empirical description of individual band profiles}\label{sub-profiles}

An observed spectral band results from an unresolved composition of multiple distinct vibrational transitions, each with its own rotational substructure, that are too close to resolve individually. At finite temperatures, each band further becomes an unresolved superposition of a large number of hot bands, separated by small anharmonic shifts, with relative weights that depend on temperature. 
Note that Doppler broadening contributes only negligible broadening for the molecules under consideration, as they are too heavy to have significant thermal velocities at the temperatures considered. The overall band profile thus results from all vibrational and rotational contributions, whose weights evolve smoothly with temperature. Our approach is to represent this profile as a superposition of analytical functions, specifically pseudo-Voigt functions (a Lorentzian-Gaussian (L/G) ratio, see Eq.~\ref{eq:pvdef}).

\begin{figure}
    \centering
    \includegraphics[height=6 cm]{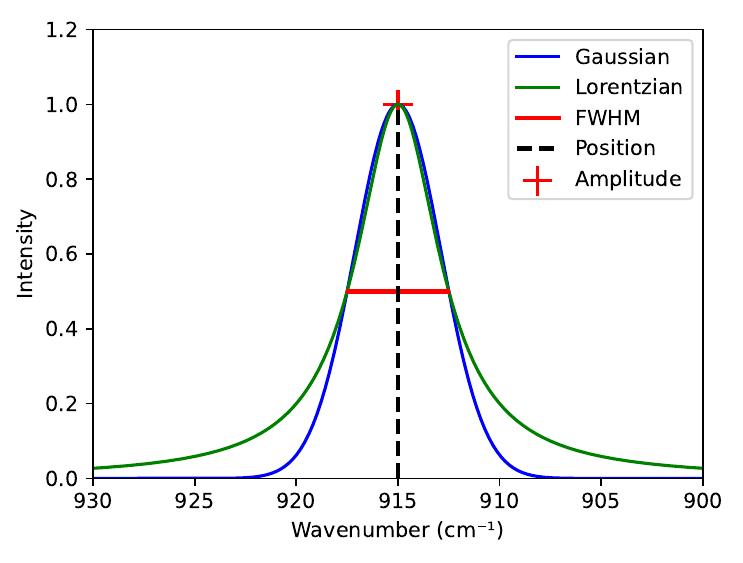}
    \includegraphics[height=6 cm]{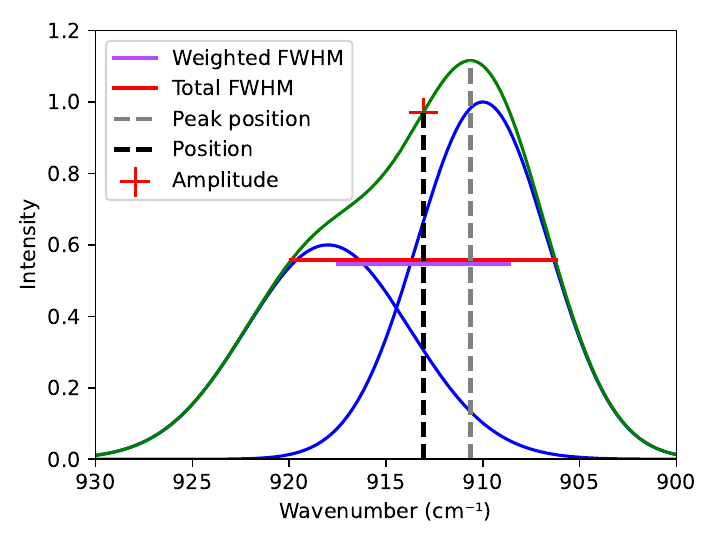}
    \caption{Definition of band parameters. Pseudo-Voigt functions (left) with limiting cases of a pure Lorentzian (L/G=1) and a pure Gaussian (L/G=0) functions. Case of merged components (right) with position and amplitude defined at half of the FWHM, and weighted FWHM defined as the average FWHM weighted by the component areas.}
    \label{fig:lump}
\end{figure} 
\vspace{0.3 cm}

When analysing a given band profile with a superposition of pseudo-Voigt functions, single components are used as much as possible. However, this representation can depend on temperature. As temperature increases, two distinct components may merge, leading to an ambiguous or ill-defined assignment of the position and width of the resulting band.
Several approaches exist, such as using the maximum of the band as the position and a weighted sum of the FWHM of the components. However, this can lead to an irregular behaviour of the band position and width with temperature. The approach chosen here, which yields smoother results, is to use the FWHM of the reconstructed band (sum of blended components) and the midpoint of this FWHM as the position. The differences are illustrated in Figure~\ref{fig:lump}. 

In experimental data, rotational broadening may be partially resolved for the out-of-plane bending bands, as shown in Figure~\ref{fig:mfitanalysis} (right panel). In this case, the PQR structure can be represented in the fit by including two additional pseudo-Voigt components to the red and blue of the central component (example on Figure~\ref{fig:mfitanalysis}). As discussed in Section~\ref{sec-tool}, this approach is simplistic. 
A more sophisticated method involves using a dedicated tool like \textsc{Pgopher} to convolve the given band with a rotational envelope, simulated using the appropriate rotational constants (obtained via theoretical calculations) at the given experimental temperature. When rotational profiles are only partially resolved, they vary little between different bands, so a single profile can serve as a proxy for all of them \citep[see][]{demyk2026}. 
For example the rotational broadening obtained with Pgopher is  8 $\pm$ 0.1 cm$^{-1}$ (distance between the maximum of P and R branches)  for all the bands of 6H-pyrene at 523~K.

\subsection{The {\rm{cosmicPAHmfit}} fitting tool}\label{sec-tool}

As mentioned previously, bands are fitted with a linear combination of pseudo-Voigt functions $\operatorname{PV}(x)$, defined as follows:  
\begin{equation} \label{eq:pvdef}
\operatorname{PV}(x) = A \left(\frac{\eta}{1 + 4\left( \displaystyle \frac{x - x_0}{f} \right)^2} + (1-\eta)\exp{\left(-4 \ln{2}\left( \frac{x-x_0}{f} \right)^2 \right)} \right)
\end{equation}
The parameters required to fully define a pseudo-Voigt function are:
\begin{itemize}
        \item the amplitude $A$,
    \item the full width at half maximum (FWHM) $f$,
    \item the position $x_0$ in cm$^{-1}$,
    \item the Lorentzian/Gaussian mixture coefficient $\eta$.
\end{itemize}

 The tool includes the possibility to use pseudo-Voigt functions with an asymmetry parameter. The area and total FWHM are unchanged by this operation, whereas the asymmetry leads to a width of (1-Asym)*FWHM and (1+Asym)*FWHM on the blue and red side of the band, respectively.
The area is calculated using the following formula:
\begin{equation} \label{eq:area}
\operatorname{Area}(x) = \frac{A \times f}{2} \left(\eta\times \pi + (1-\eta)\times\sqrt{\frac{\pi}{ln(2)}} \right)
\end{equation}

These parameters, multiplied by the number of used components, are fitted to obtain the analytical decomposition of a band.

Since bands are assumed to evolve smoothly with temperature, we developed \texttt{cosmicPAHmfit}, a tool that streamlines the fitting process across all temperatures. \texttt{cosmicPAHmfit} is implemented in Python 3.12 using Tkinter for the interface. Some parameters may be fixed or constrained to vary within a limited range. The code performs a non-linear least-squares fit using the Levenberg-Marquardt method available in the lmfit library. The lmfit implementation allows to apply constraints on the fitted parameters, and we use this to always enforce positive amplitudes for all components. For the lowest temperature, it uses an initial guess provided by the user. For each subsequent higher temperature, the code uses the results from the previous temperature as the new initial guess and determines the new fit iteratively refining them. The user must monitor and validate each fit or intervene to correct it if problems arise. This approach naturally ensures a smooth evolution of each parameter with temperature. 

The derived band positions and widths are then displayed as a function of temperature, and their evolution is adjusted using simple analytical functions, such as low-order polynomials or collections of broken straight lines. This process yields a set of \emph{empirical anharmonicity factors}. 

Here, we discuss as typical practical examples the cases of the decomposition of the 11.9\,$\mu$m band of pyrene and the more complex case of the 3\,$\mu$m band of 6H-pyrene. To be analysed properly, the data are assumed to be prepared beforehand: any underlying baseline (continuum) must be removed so that they only contain the bands to decompose.

The 11.9\,$\mu$m band of pyrene is a simple case of the CH out-of-plane bending modes, which can be reasonably well fitted with one component (Figure~\ref{fig:comppy}). A decomposition in 3 components improves the fit and is used for a simplistic removal of the rotational width. The difference between the central component width and the one-component fit width provides an approximation of the rotational broadening of the pyrene-like species, an approach followed by \cite{demyk2026}. There is, however, evidence that the blue and red components are more than just the R and P branches of the fundamental CH out-of-plane bending mode.  Anharmonic calculations indeed reveal vibrational transitions on both the red and the blue sides (see Figure~3 in \cite{chakraborty2021}; see \href{https://cosmicpah-irdb.irap.omp.eu/science/Pyrene/Gas/26/sample/?overplot-tab\#/}{Id~26} in \texttt{CosmicPAH\textendash IRDB} database).  Nevertheless, the simplistic approach still provides values for the rotational broadening that agree with theoretical predictions \citep[see][]{demyk2026}.

    \begin{figure*}
    \centering
    
    \includegraphics[width=1\linewidth]{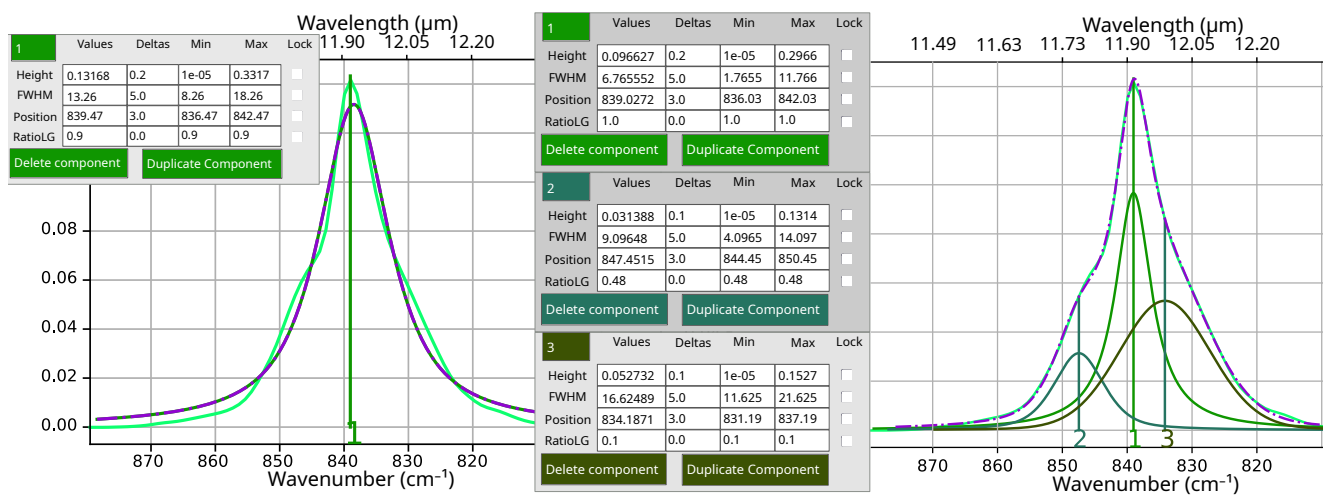}
    \caption{\texttt{cosmicPAHmfit} tool - analysis window. Decomposition of the 840~cm$^{-1}$ (11.9~$\mu$m) band of pyrene using either a single component (left) or three components (right).}
    \label{fig:mfitanalysis}
\end{figure*}

\begin{figure}
     \includegraphics[height=7.1 cm]{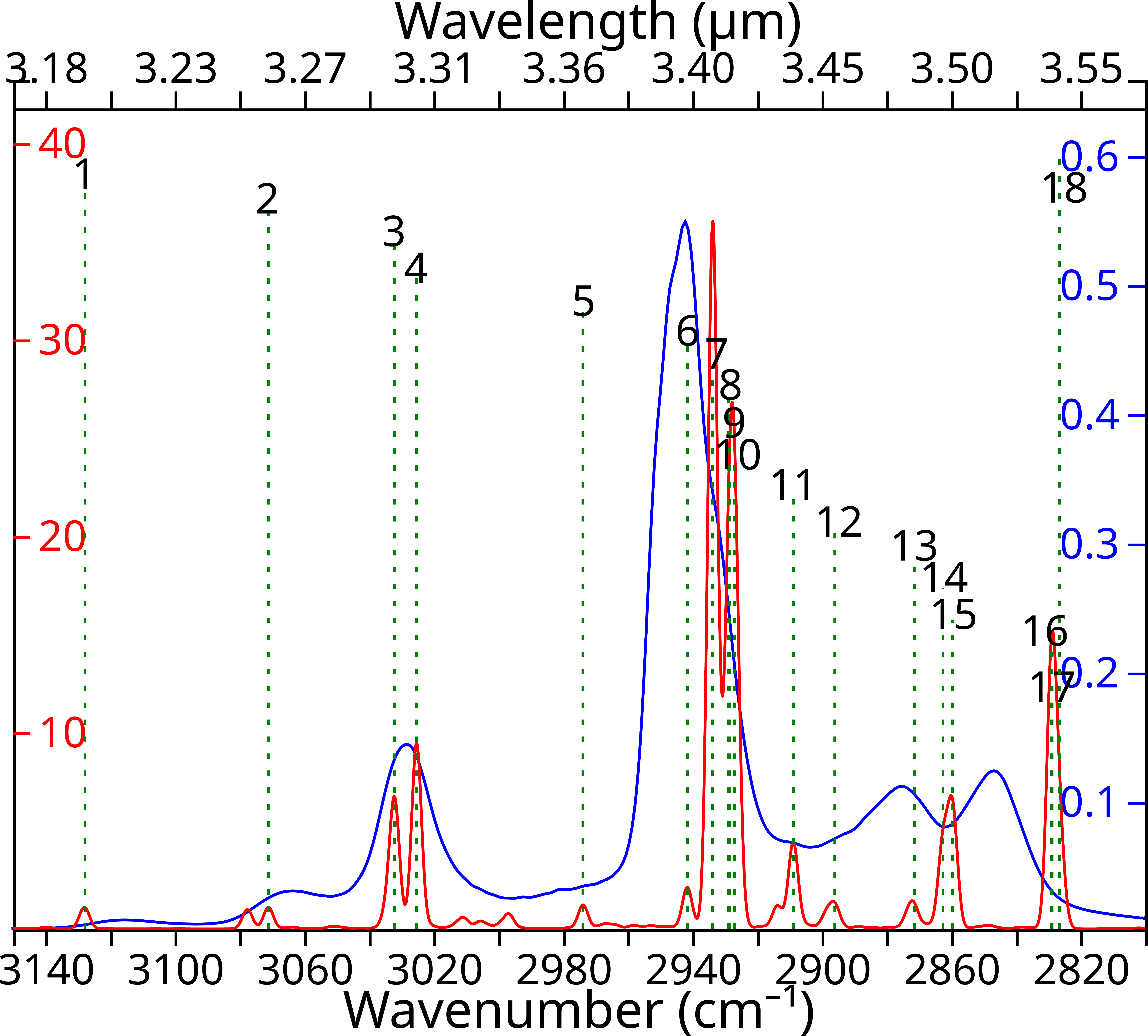}
     \includegraphics[height=6.3cm]{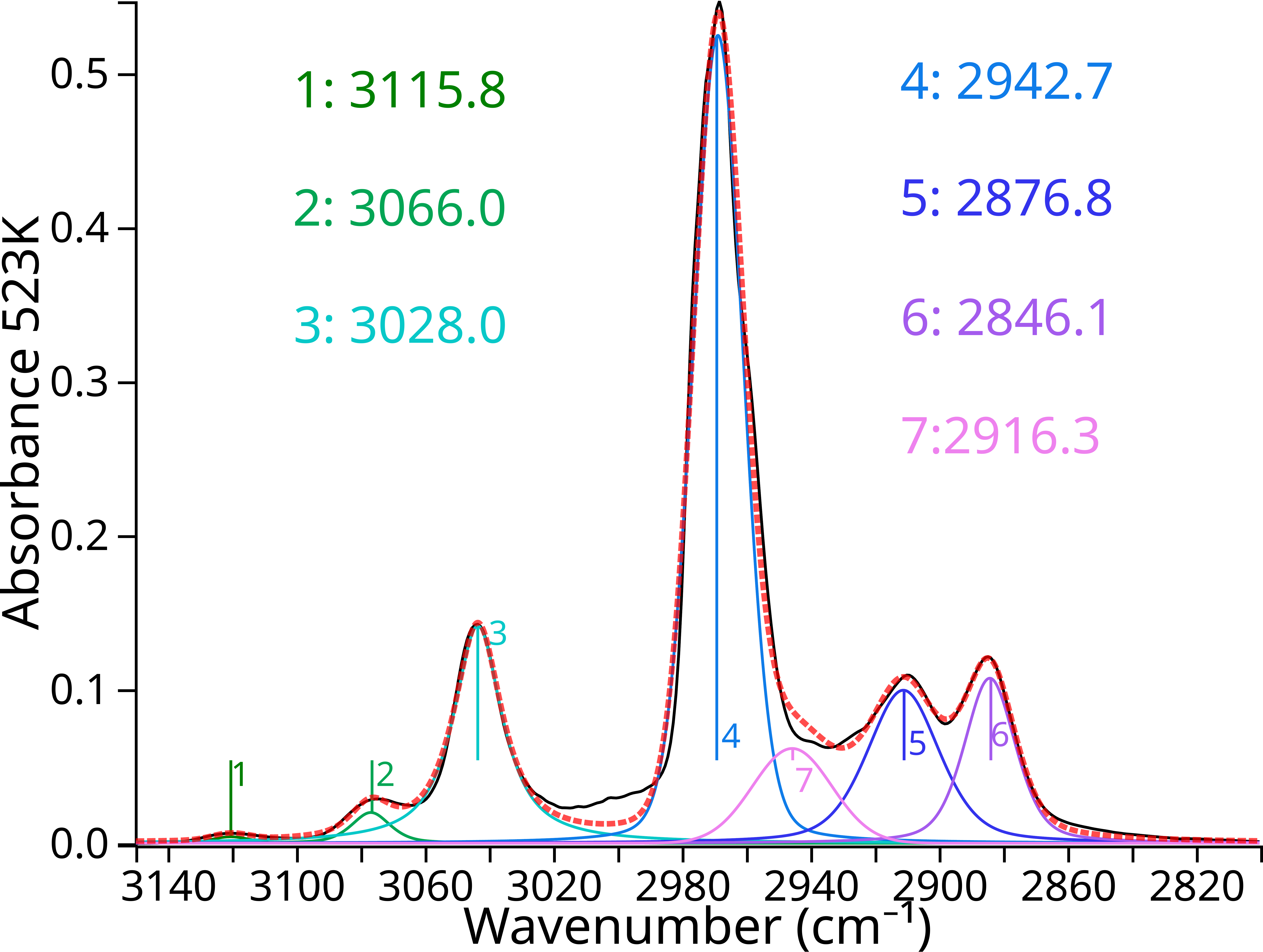}
    \caption{IR spectrum of 6H-pyrene in the 3~$\mu$m range. Left: experimental gas-phase spectrum at 523~K compared with the anharmonic spectrum at 0~K calculated with AnharmoniCaOs following \cite{demyk2026}.
    Right: decomposition of the gas-phase spectrum using \texttt{cosmicPAHmfit}. The positions of the seven components are listed in cm$^{-1}$. For an interactive use, the figures can be accessed on the \texttt{CosmicPAH\textendash IRDB} website using the following link: \href{https://cosmicpah-irdb.irap.omp.eu/science/1,2,3,6,7,8-hexahydropyrene/Gas/4/sample/?overplot-tab\#/3/5/5/3/3150/2800/800/600/523/-1}{Left graph} \href{https://cosmicpah-irdb.irap.omp.eu/science/1,2,3,6,7,8-hexahydropyrene/Gas/4/spectrum/profiles/3030}{Right graph}}
    \label{fig:overplot_H6Pyr}
\end{figure}

\begin{figure*}
    \centering
    \includegraphics[width=1\linewidth]{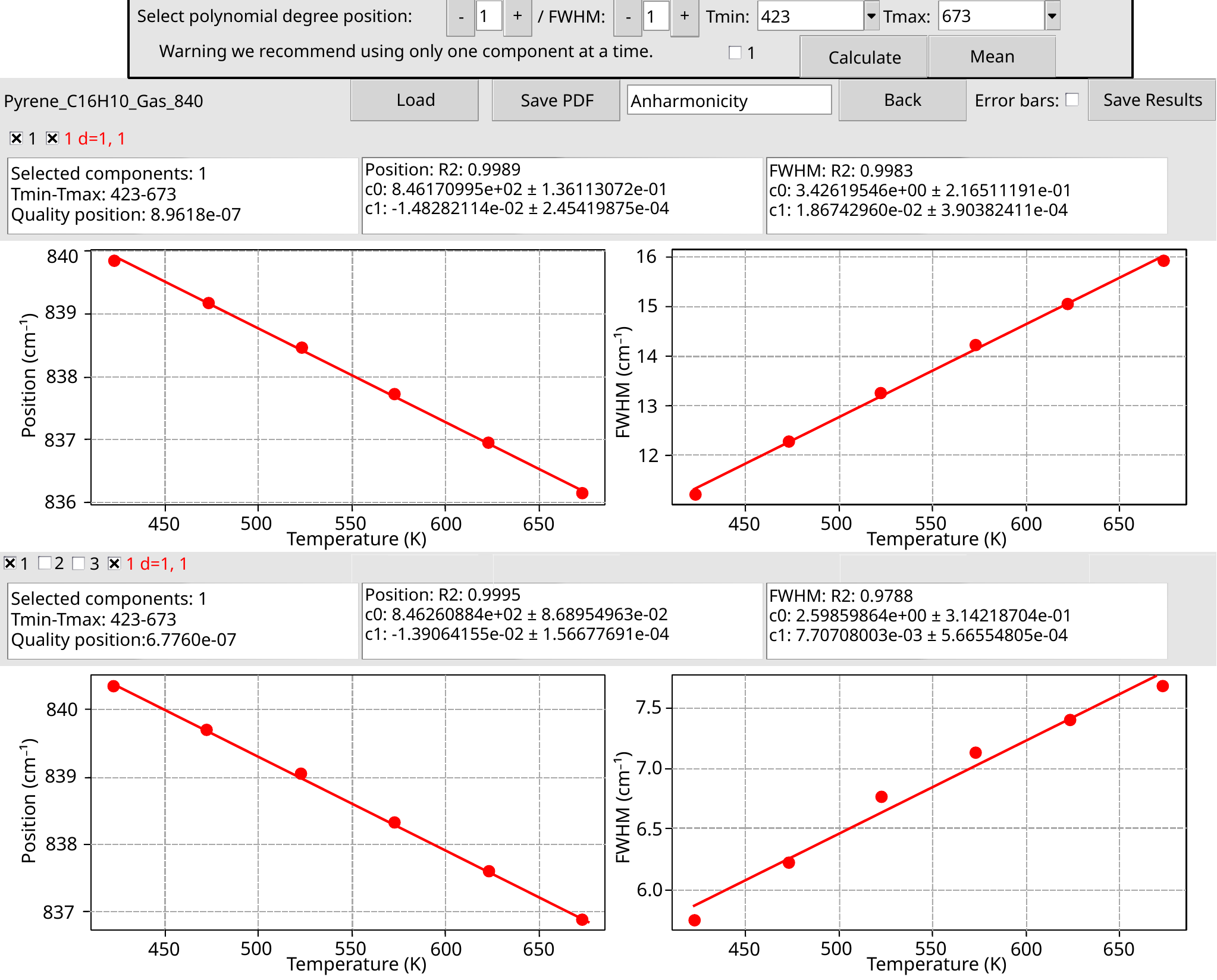}
    \caption{\texttt{cosmicPAHmfit} tool - anharmonicity window.  Empirical anharmonicity factors for the 840~cm$^{-1}$ (11.9~$\mu$m) band of pyrene. A linear function  $c0+c1*T$ is used for the fit. The results correspond to the single component in Figure~\ref{fig:mfitanalysis} (top panel) and to the central component of the three components in Figure~\ref{fig:mfitanalysis} (bottom panel). }
\label{fig:mfitanharm}   
\end{figure*}  

A more complex case involves the CH stretching modes of 6H-pyrene in the 3.3-- 3.5\,$\mu$m range, for which seven components are identified (Fig.~\ref{fig:overplot_H6Pyr}), some of which are unresolved (e.g., component 7). In such cases, the user must choose a reasonable number of components, guided by harmonic or anharmonic spectra, and define a correct range of variation for parameters at each iteration. 

Even though the tool helps streamline the procedure, following the evolution of spectra with a correct decomposition for each temperature requires physical understanding: the components must maintain coherence across temperatures, and effects such as intensity transfer or component swapping must be avoided. 

The results of the decomposition are stored in a first file, which is used for the next step. This file contains the parameters of the components at all temperatures as well as the names of the files of the spectra used for the decomposition. Once the decomposition at all temperatures has been satisfactorily achieved, the application assists with the second step: the extraction of empirical anharmonicity factors for position and FWHM. Using the first result file, the code performs secondary fits on the temperature evolution of the fitted parameters. The temperature range and order of the polynomial can be chosen to extract the anharmonicity factors. As an example, Figure~\ref{fig:mfitanharm} shows the results obtained for the 11.9\,$\mu$m band of pyrene when using either a single component or three components in the decomposition (Figure~\ref{fig:mfitanalysis}). The latter case corresponds to the scenario where the contribution from rotation is removed \citep{demyk2026}. In all cases, a single linear fit is sufficient to describe the evolution of the position and the FWHM with temperature. The technique we used to remove rotational broadening does not significantly modify the band position but does reduce the FWHM, as expected.

Following the extraction of anharmonicity factors, a file is generated containing both the decomposition data and the extracted anharmonicity factors. The file is formatted in a way suitable for ingestion into our database, \texttt{CosmicPAH\textendash IRDB}, which is described in the next section (Section~\ref{sec-database}).

\section{The CosmicPAH\textendash IRDB web application }\label{sec-database}

In order to gather spectral information on the evolution of the IR spectra of PAHs with temperature and provide empirical anharmonicity factors for use in emission models of excited PAHs, we have developed the \texttt{CosmicPAH\textendash IRDB} web application. The web application contains both experimental and theoretical spectra, as well as the results of analyses performed using the multi-component spectral fitting tool \texttt{cosmicPAHmfit}. It  provides an overplot functionality, which is convenient for spectral analysis and band assignment. The website also supports access through machine requests.

\begin{figure}
    \centering
    \includegraphics[width=1\linewidth]{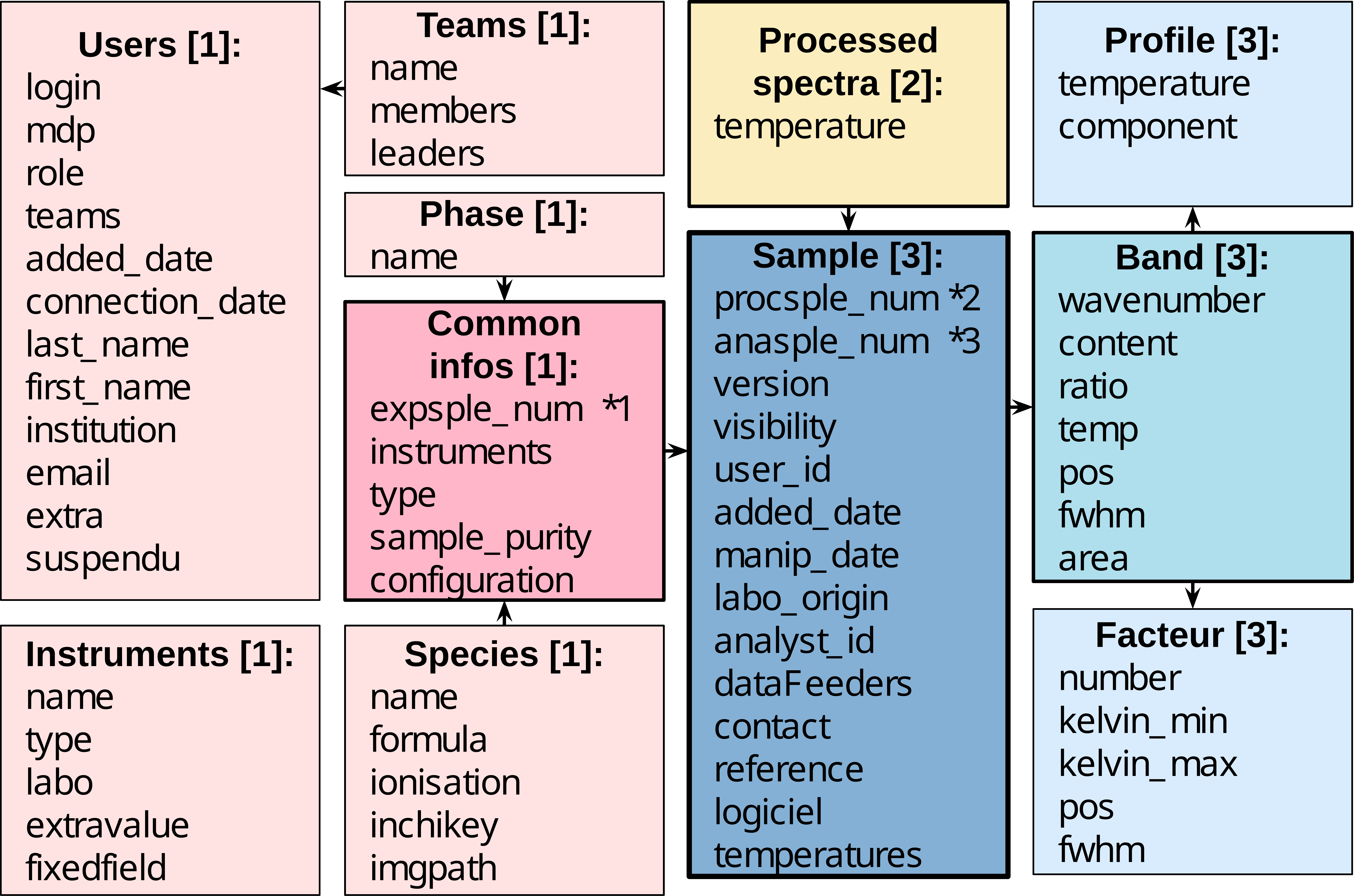}
    \caption{Structure of the \texttt{CosmicPAH\textendash IRDB} database.}
    \label{fig:webstruc}
\end{figure}

\texttt{CosmicPAH\textendash IRDB} is a Python web application. The website backend is developed in Python, while the frontend uses JavaScript with Jinja2 templating engine and the Flask framework. The schematic structure of the database is given  in Figure ~\ref{fig:webstruc}. The web application uses PostgreSQL and is deployed within a Docker container. It runs on a server that hosts data processing and valorisation services in the context of open science, specifically the \textit{Cloud Recherche Occitanie} (CROcc). \texttt{CosmicPAH\textendash IRDB} and \texttt{cosmicPAHmfit} are accessible and promoted through the \texttt{Cosmic PAH portal}, which is designed to facilitate access to databases and tools for studying PAHs and related molecular species (e.g. fullerenes) in astrophysical environments.

\begin{figure*}
    \centering
    \includegraphics[width=1\linewidth]{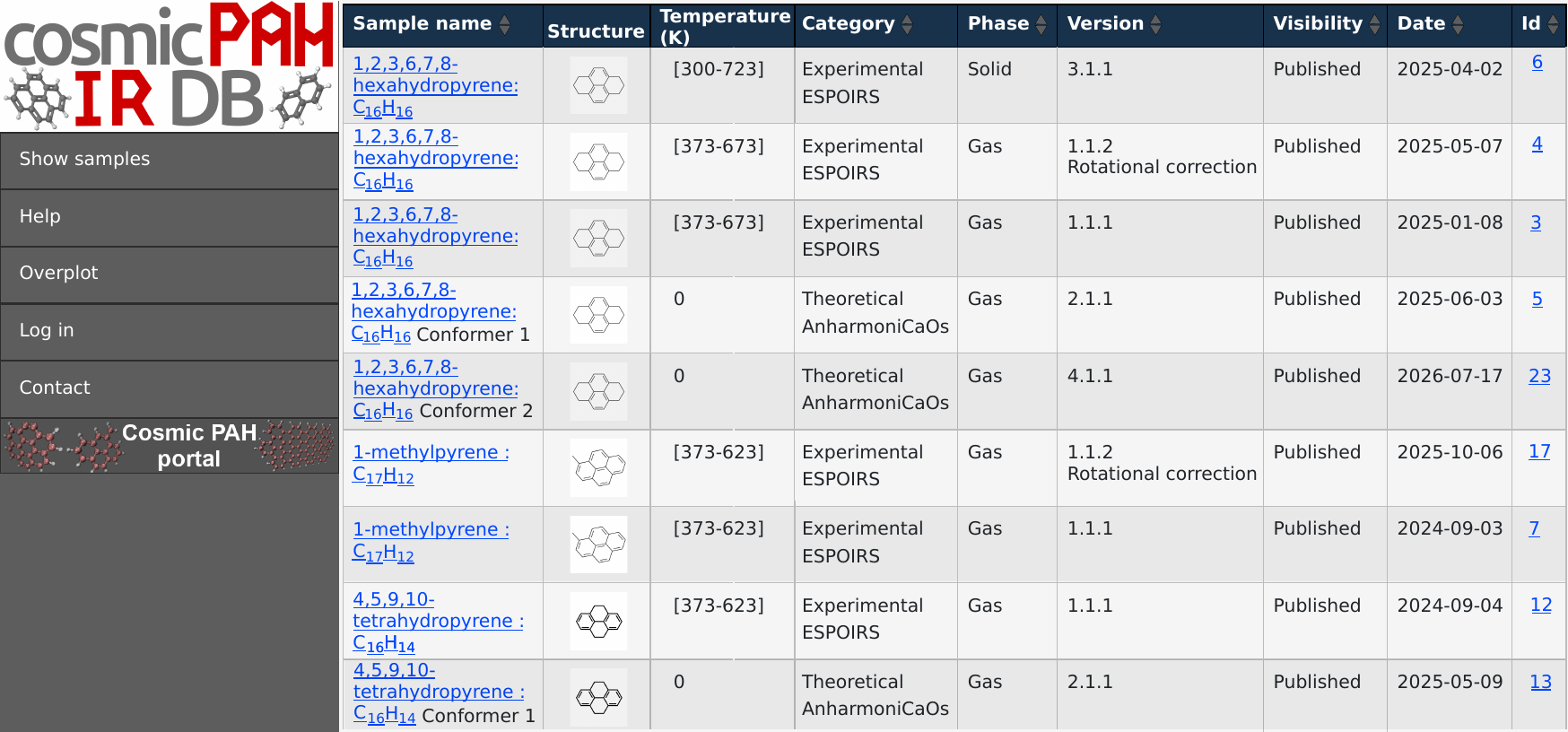}
    \caption{\texttt{CosmicPAH\textendash IRDB} website. Illustration of the content provided in \texttt{Show samples}.}
    \label{fig:mainweb}
\end{figure*}  

\subsection{Database content}
The database contains spectral information for different PAH molecules (\texttt{Show Samples} tab). Associated fields (Figure~\ref{fig:mainweb}) include a description of the molecule (name, chemical formula, and structure) and of the dataset content and type (studied temperatures, category: experimental/theoretical, and the instrument used).
Additionally, a version number, visibility flag, date, and Id are provided for each sample.

\begin{figure*}
    \centering
    \includegraphics[width=1\linewidth]{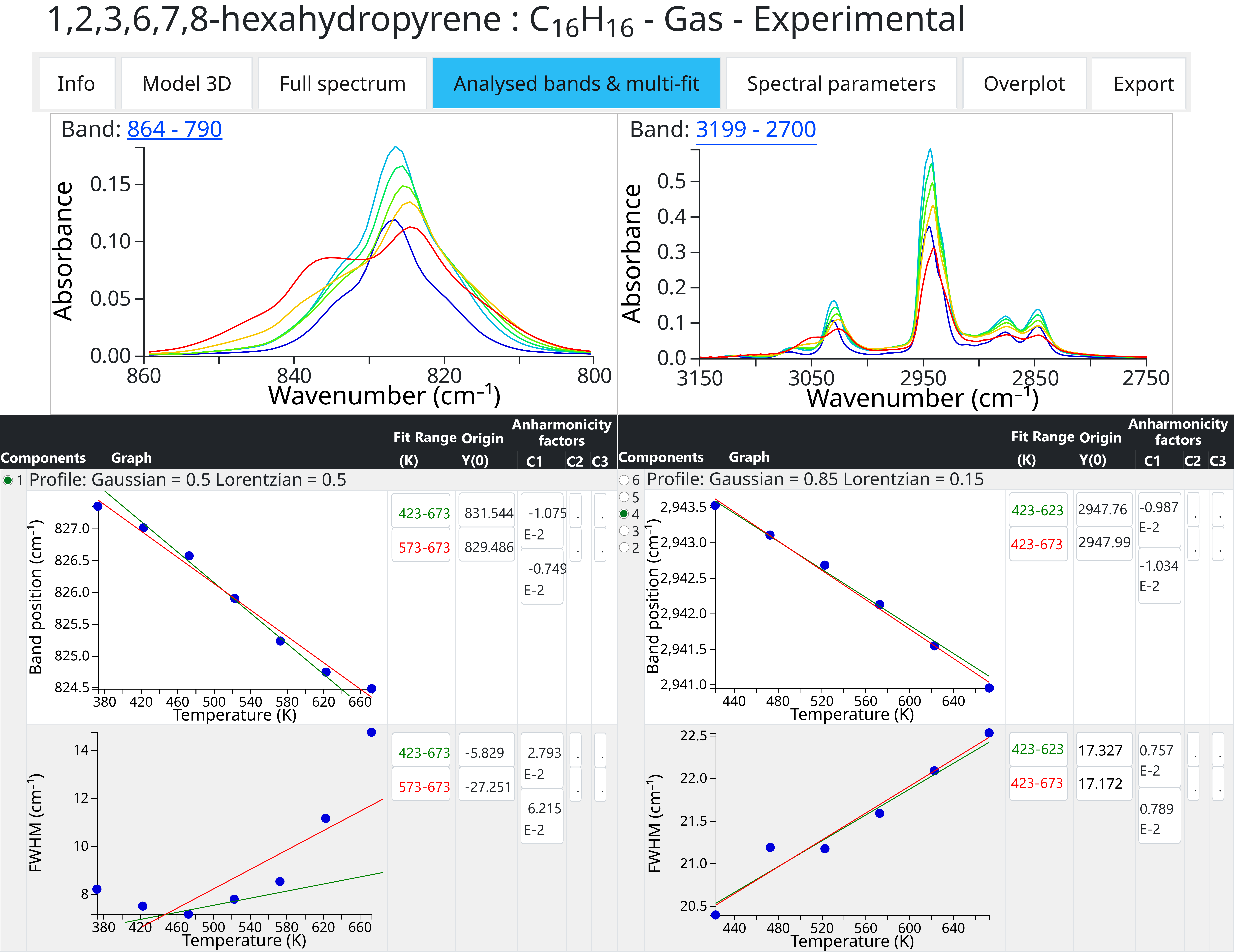}
    \caption{\texttt{CosmicPAH\textendash IRDB} website. Illustration of the content of \texttt{Analysed bands \& multi-fit} for the gas-phase spectrum of 6H-pyrene (Id~3).}
    \label{fig:bandstab}
\end{figure*}

After selecting a sample, users can access metadata (\texttt{Info}), a molecular representation (\texttt{Model 3D}) of the molecule and the spectral data. Graphs of the full spectra at all available temperature can be interactively zoomed in (\texttt{Full spectrum}). \texttt{Analysed bands \& multi-fit} provides access to the detailed band analysis performed using the \texttt{cosmicPAHmfit} tool (Figure~\ref{fig:bandstab}), exhibiting the spectral decomposition and the derived anharmonicity factors. Individual components used in the band analysis can be interactively selected. \texttt{Spectral Parameters} include Tables summarizing the spectral parameters derived -- band positions, widths, and intensities at different temperatures. Note that spectral analysis is not available for discretised histogram spectra, which must first be convolved with a custom profile.  \texttt{Export} allows users to extract all spectral data, parameters related to their analysis, and metadata. The structure of the downloaded files is also the one required for automated ingestion of new samples into the database.

To distinguish between different analyses of the same species, a versioning system has been implemented. The versions follow an \texttt{X.Y.Z} format, structured as follows:
\begin{itemize}
\item \texttt{X}: Incremented when a new experimental or theoretical measurement/calculation is added.
For example: \texttt{1.1.1} could represent an experimental measurement with ESPOIRS and \texttt{2.1.1} could represent a theoretical calculation of the same species using AnharmoniCaOs.
\item \texttt{Y}: Incremented when a new post-processing of the data is applied, such as continuum subtraction for experimental spectra or convolution of the IR transitions to generate theoretical spectra.
\item \texttt{Z}: Incremented when a new analysis of the same processed data is performed, such as a different decomposition or the application of rotational corrections.
\end{itemize}

At the time of this article, 21 datasets have been made public in relation with the study of pyrene (C$_{16}$H$_{10}$) and its methylated (C$_{17}$H$_{12}$) and hydrogenated derivatives (C$_{16}$H$_{16}$, C$_{16}$H$_{14}$, C$_{16}$H$_{12}$,  \cite{demyk2026}). In addition to our own production, the database is designed to accommodate datasets from other providers. Nine datasets corresponding to infrared multiphoton dissociation spectra of cationic PAHs obtained at FELIX \citep{oomens2000, oomens2001, oomens2001_jpca, banisaukas2003,oomens2006} are readily accessible.

\subsection{Overplot functionality}
In addition to data content, the database offers an overplot functionality to superimpose two spectra. This is particularly useful for comparing, for example, experimental and theoretical spectra. For theoretical spectra at 0~K, labels for the transitions can be displayed. A band intensity threshold should be set to avoid overloading the page (default value: 0.5~km~mol$^{-1}$). The labels are displayed on the graph, and the first table is populated with the labels, positions, and intensities of the transitions. In the case of AnharmoniCaOs data, the table also includes detailed transition assignments. The labels and table can be hidden by unchecking the corresponding option. 

\texttt{Overplot} is accessible from all samples included on the database (Figure~\ref{fig:bandstab}) to support spectral analysis (see, eg., Figures~\ref{fig:comppy} and \ref{fig:overplot_H6Pyr}). It is also accessible from the general menu on the left (Figure~\ref{fig:mainweb}). The latter extends the possibility to overplot two spectra, which could be from outside the database, including spectra accessible locally by the user. It therefore serves as a functionality independent of the database content. When using \texttt{Overplot} through a sample, the first spectrum in the superimposition is the one associated with the selected sample, whether experimental or theoretical. When using \texttt{Overplot} from the left panel, both spectra must be selected.

To find datasets, \texttt{Find URL from...} feature allows users to search for external dataset URL links by molecular formula. Currently, the NASA Ames Spectroscopy database and the Theoretical spectral database of Polycyclic Aromatic Hydrocarbons can be queried. The retrieved URL must then be entered into the dedicated URL field. Alternatively, the \texttt{Find Id} feature can search for the Id of a spectrum available on \texttt{CosmicPAH\textendash IRDB}. In this case, either the Id or the URL of the dataset can be used, and a specific temperature can be selected from the displayed options. For more details refer to the \href{https://cosmicpah-irdb.irap.omp.eu/help}{help page}. Finally, users can input their own datasets using the \texttt{Use local...} option. With this option, no data is uploaded online; it remains stored locally on the user's computer.
If the queried dataset includes a list of IR vibrational transitions, these transitions are retrieved, their labels are displayed, and they are convolved using the FWHM parameter.

\subsection{Data finding/retrieving and accessibility}\label{sub-use}

Users can navigate the website to select and download data of interest in a zip file. The website also supports data access via URL requests. The dedicated format .../machine/<Molecular Formula>/<Phase>/<Category>/0 retrieves the list of available data.
If any \texttt{<field>} is set to \texttt{0}, the full list is displayed. For example:\\
\url{https://cosmicpah-irdb.irap.omp.eu/machine/0/Gas/0/0}
displays the list of experimental and theoretical data for all species in the gas phase.
The data can then be extracted in a convenient format by using the same URL but replacing the \texttt{0} at the end with an \texttt{Id}. The result is a JSON output containing all data of the queried sample.
For convenience, all figures generated from the \texttt{CosmicPAH\textendash IRDB} website or the \texttt{cosmicPAHmfit} application can be regenerated using result files downloaded from the website. The fits can also be retried starting from the parameters of the result files in the application.\\

Both \texttt{CosmicPAH\textendash IRDB} and \texttt{cosmicPAHmfit} are developed within the context of the \textit{Observatoire Virtuel du Grand Sud-Ouest (OVGSO)}. When using these tools, please cite this article as well as the relevant references listed on the main page of the website. Additionally, each sample in the database includes a \texttt{DOI/Reference} field containing links to the original published data, which must also be cited. \texttt{CosmicPAH\textendash IRDB} is licensed under CC BY-SA 4.0, and \texttt{cosmicPAHmfit} is licensed under GPLv3.\\

\section{Application to UV-excited PAHs}\label{sec-application}

The most complete spectral datasets on the IR emission spectra of highly-excited PAHs have been obtained by the group of R. Saykally at Berkley University \citep{schlemmer1994, cook1996}. In these experiments, PAHs gain energy upon irradiation by an excimer laser at 248~nm (5~eV) following laser-induced desorption. The hot PAH plume is viewed by the IR detector for a relatively short time ($\sim$100~$\mu$s). These short time differs from the long times available in interstellar environments and this can affect the observed IR emission band profiles. However these spectra are the only ones available for high-excitation conditions and it is worth to compare them with the results of models (cf. \cite{chen2018AandA} in the case of anharmonic calculations).

\begin{figure*}
    \centering
    \includegraphics[height=44. mm]{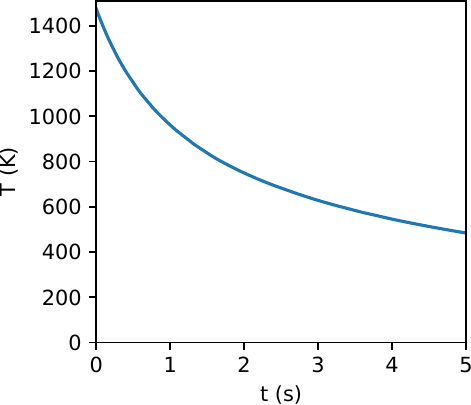}
    \includegraphics[height=44. mm]{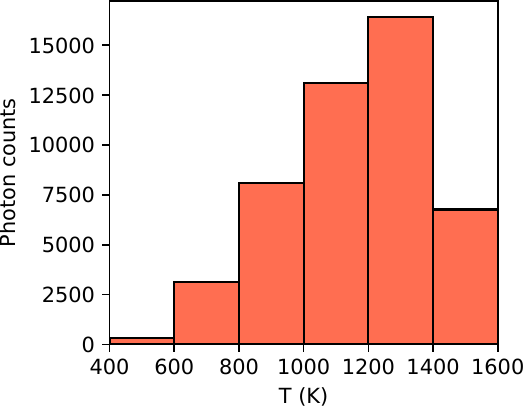}
    \includegraphics[height=44. mm]{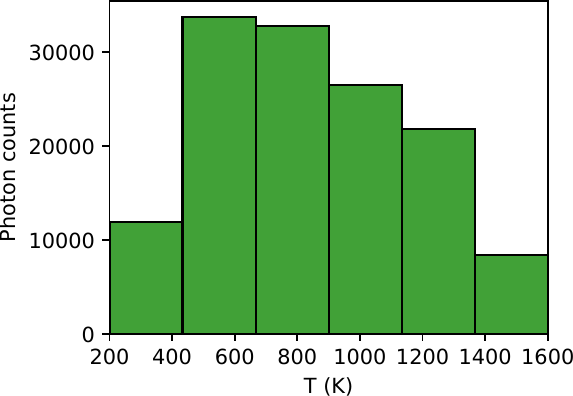}
    \caption{Results from the Monte Carlo code. Cooling curves of pyrene following the absorption of an energy of 5~eV (left). Temperature distribution of the IR photons emitted by pyrene in the 3.3~$\mu$m (middle) and 11.9~$\mu$m (right) bands following the absorption of an energy of 5~eV.}
    \label{fig:model}
\end{figure*}

\begin{figure}
    \centering
    \begin{subfigure}[b]{0.48\textwidth}
        \includegraphics[width=\textwidth]{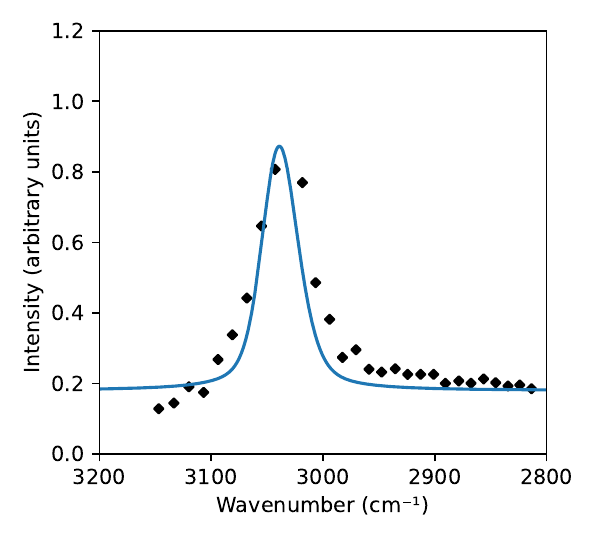}
    \end{subfigure}
    \hfill
    \begin{subfigure}[b]{0.48\textwidth}
        \includegraphics[width=\textwidth]{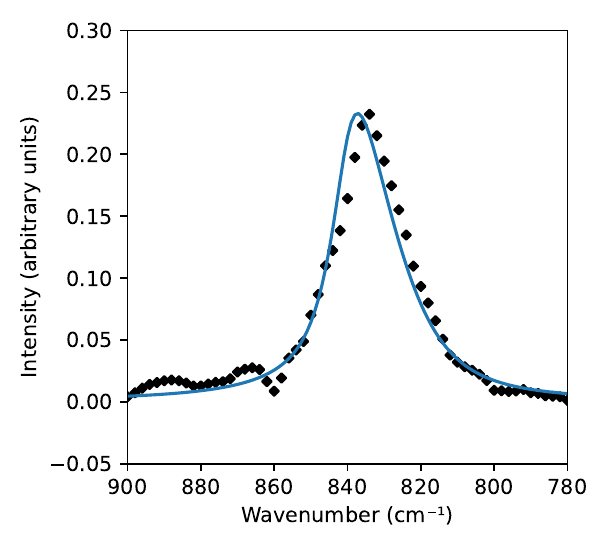}
    \end{subfigure}
\caption{Comparison of the band profiles measured from UV-excited PAHs (results from R. Saykally's laboratory) with simulated spectra using our Monte Carlo code and the spectral parameters of the \texttt{CosmicPAH\textendash IRDB} (dataset \href{https://cosmicpah-irdb.irap.omp.eu/science/Pyrene/Gas/2/sample/}{Id~2}). The laboratory data were scanned from the original articles: \cite{wagner2000} for the 3.3~$\mu$m band and \cite{cook1996} for the 12~$\mu$m band.}
    \label{fig:emission_bands}
\end{figure}

As a proof of principle, we used the following methodology to simulate spectra that can be compared to these spectra. First, we use a Monte Carlo code (improved version of the one discussed in \cite{joblin2002}) to calculate the cooling curve following the initial absorbed energy (5~eV in this case; Figure~\ref{fig:model}) and to record the emission of IR photons in a given IR band as a function of temperature (Figure~\ref{fig:model}). Here, for pyrene, we rely on the absolute intensities previously determined in the gas phase \citep{joblin1995, mulas2018}. Since the Monte Carlo code uses the list of harmonic frequencies to calculate the density of states, we selected experimental band intensities that can be associated with harmonic modes. This excludes the five combination bands between 1650 and 1900~cm$^{-1}$.

The spectral profiles of the bands of interest are then reconstructed using the spectral parameters listed in the \texttt{CosmicPAH\textendash IRDB} database (empirical laws for the band positions and widths and L/G ratio for the corresponding pseudo-Voigt profile). The input file of the Monte Carlo code consists in the list of modes and IR intensities. The file was constructed by combing the list of harmonic frequencies from the \href{https://cosmicpah-qcals.oa-cagliari.inaf.it/}{Theoretical spectral database of polycyclic aromatic hydrocarbons} with the absolute intensities derived from experimental data \citep{joblin1995} and listed in Table~IV of \cite{mulas2018}. Only the main bands corresponding to fundamentals have been considered. Detailed spectral profiles were calculated only for the two most intense bands in the 3 and 12~$\mu$m range for comparison purposes with UV-excited spectra. We selected dataset \href{https://cosmicpah-irdb.irap.omp.eu/science/Pyrene/Gas/2/sample/}{Id~2} in the \texttt{CosmicPAH\textendash IRDB} database, for which a simple analysis was performed with a single component including the contribution of rotation.

The comparison of the simulated spectra with the experimental emission spectra following UV excitation  is reported in Figure~\ref{fig:emission_bands}. The agreement appears very good, which supports our approach and the associated simplifications, including the fact that we extrapolate spectral parameters derived in a temperature range below 700~K to temperatures as high as 1600~K, which can easily be reached upon absorption of a UV photon. The observed slight blue shift of the simulated spectra might be due to the fact that our simulations extend to a much longer time relative to the experimental timescale. In addition, there is an emission plateau to the red of the 3.3~$\mu$m band that is not included in our simulations. This plateau is likely due to numerous combination bands.

\section{Discussion}\label{sec-discussion}

Due to the complexity of the IR spectra of these large molecules, performing spectral decomposition is delicate and involve subjective choices in the selection of the number of components and their associated parameters in the spectral fitting.
Whereas the \texttt{cosmicPAHmfit} tool supports the user, it cannot handle this problem. It is important to point out the difficulty to constrain the fit and the variability of the results for different sets of parameters. Even in the relatively simple case of the CH out-of-plane band (e.g., the 840~cm$^{-1}$ band of pyrene), there is a rotational band profile in the case of thermal equilibrium that comes on the top of weak unresolved hot bands on band wings (Figure \ref{fig:mfitanalysis}). A spectroscopist's expertise helps to optimise the use of the provided tool, but still subjectivity in choices can lead to somewhat different results.

A number of options have been implemented in \texttt{cosmicPAHmfit}, including the ability to use an asymmetry parameter on the Voigt profiles \citep{stancik2008} and to allow the L/G ratio to evolve with temperature. However, while this results in only a marginal improvement in the quality of the fits, it would require implementing asymmetric profiles and temperature-dependent L/G ratios in the astronomical modelling code that uses these data. Given that the impact on the final astronomical models would be significant, we have, for the time being, chosen not to use asymmetric profiles and to fix the L/G ratio, keeping the modelling code simpler.
In general, we recommend -- based on our experience --  minimizing the number of parameters by using the minimum number of symmetric components, following Occam’s razor’s criterion. 
Even if a band is resolved into two close components at the lowest temperatures, it can be better to use a single component to follow the band evolution with temperature.

Another option concerns the derivation of empirical anharmonicity factors, which can use polynomials rather than linear fits. As illustrated in \cite{chakraborty2019}, there are significant changes in linear trends when extending the range of studied temperatures. One can therefore derive broken lines to extract anharmonicity factors over a number of limited temperature ranges. Alternatively, a single polynomial function can also be used. However, whatever the degree of the fit, one should be very careful when extrapolating the data outside the measured temperature range. Our recommendation is to use a linear extrapolation toward high temperatures, an approach often required in astrophysical applications where PAHs are UV-excited and reach temperatures in the 1000--2000~K range (e.g., Figure~\ref{fig:model}). An exception to this recommendation concerns bands whose FWHM decreases with temperature. In these rare cases, we limit the evolution of the width at the maximum temperature measured, for their use in our simulations of emitted spectra.

At this stage, \texttt {CosmicPAH\textendash IRDB} accepts data of different types, including tabulated experimental and theoretical spectra, as well as calculated vibrational transition files.  So far we have made public only a limited number of datasets. Other samples of thermally-excited PAHs (measured with ESPOIRS setup) will be released by our group after analysis and publication. Also new datasets will be added to those already available such as deMonNano dynamic molecular calculations or FELIX IRMPD measurements. \texttt{CosmicPAH\textendash IRDB} is open to collaboration and contributions, with the goal of improving dataset accessibility, while respecting data ownership.  To contribute to the database, please refer to the contact page on the website. Different user profiles are available to manage data access. Additionally, to facilitate collaborating projects, teams can be created. 

We have shown how theoretical spectra can guide the analysis process of experimental data. However, the goal is also to benchmark theoretical methods by comparison with experimental results \citep{mulas2018,lemmens2021, chakraborty2021}. The overplot functionality facilitates the comparison between spectra. Additionally, theoretical data can be analysed using \texttt{cosmicPAHmfit}, and the derived empirical anharmonicity factors can be compared to those obtained from experimental datasets. As of now, we have analysed the AnharmoniCaOs dataset of pyrene up to a temperature of 523~K. The overall spectral agreement between the theoretical and experimental spectra at 523~K is good in the 12--14~$\mu$m region and less satisfactory in the 3~$\mu$m region (Figure~\ref{fig:comppy}). For theoretical data, the empirical anharmonicity factor, which quantifies the slope of the linear evolution of the band position with temperature, is lower by a factor 2 and 3--5 in the 12--14~$\mu$m and 3.3~$\mu$m ranges, respectively (see \href{https://cosmicpah-irdb.irap.omp.eu/science/Pyrene/Gas/41/sample/}{Id~41} compared with \href{https://cosmicpah-irdb.irap.omp.eu/science/Pyrene/Gas/22/sample/}{Id~22}). This can be at least partially explained by the challenges that theoretical methods face in C--H stretch range and illustrates that further work is required before using theoretical methods to systematically obtain data on the evolution of IR spectra with temperature.

Their use as an alternative for laboratory measurements would enable the construction of large databases of spectra and derived empirical anharmonicity factors. Indeed, systematic laboratory measurements are challenging due to the large effort required to measure a large number of species. In addition, they face several limitations in exploring all parameters, including the chemical availability of species, charge state (as absorbance measurements on gas-phase charged PAHs have not yet been achieved), temperature range (ideally, data should be collected from 0~K to the dissociation limit), and out-of-thermal-equilibrium situations (e.g., molecules are vibrationally but not rotationally hot in astrophysical environments).

The empirical anharmonicity laws we generate for band position and FWHM can be interfaced with PAH emission models for astrophysical applications. Integrated band intensities are also critical for these models, as they influence the cooling curve and the number of IR photons emitted in a given band (Figure~\ref{fig:model}). Intensity values are retrieved from the \texttt{cosmicPAHmfit} analysis and are tabulated in \texttt {CosmicPAH\textendash IRDB}. For ESPOIRS data, only absorbances are obtained, and a scaling factor is required to convert them into absolute values per molecule \citep{demyk2026}. In the case of pyrene, we rely on absolute intensities previously determined in the gas phase (see Section~\ref{sec-application}). 

Our primary motivation in analysing sample \href{https://cosmicpah-irdb.irap.omp.eu/science/Pyrene/Gas/2/sample/}{Id~2} was to derive empirical anharmonicity factors for the two strongest bands, enabling comparison with UV-excited spectra (Figure~\ref{fig:emission_bands}). A systematic analysis of all bands would require further work and was not considered at this stage.

Astrophysical applications are the main motivation for this work. A \texttt{Show simulations} tab is therefore included in \texttt{CosmicPAH\textendash IRDB}, where calculated band profiles for specific UV-excited PAHs will be made available. Future developments could include providing grids of emission spectra for a given species at different internal energies. These spectra can then be used as input in astronomical models, which provide physical and chemical conditions (e.g., the UV radiation field), to calculate AIB spectra in astrophysically relevant environments. 

\section{Conclusions and perspectives}\label{sec-conclusions}
In this work, we have developed a methodology for quantifying the effect of temperature on the IR band profiles of PAHs and providing empirical anharmonicity factors that can be used to generate simulated spectra of UV-excited PAHs.  Comparison with laboratory spectral data on UV-excited pyrene (Section~\ref{sec-application}) demonstrated the potential of this approach in the simulation of the emission spectra of PAHs in astrophysical environments.

The methodology requires a dataset of IR spectra at different temperatures and relies on the multi-component spectral fitting tool \texttt{cosmicPAHmfit}. Our analysis of experimental spectra (pyrene and 6H-pyrene) led us to implement specific solutions, for instance, on how to handle spectral complexity and band merging and in the choice of the empirical anharmonicity functions to quantify the evolution of band position and FWHM with temperature. The methodology can indifferently consider experimental and theoretical datasets when available for a range of temperatures.

The \texttt {CosmicPAH\textendash IRDB} web application has been developed alongside the \texttt{cosmicPAHmfit} fitting tool to gather spectra and the results of their spectral analysis, including the implementation of different scenarios. The web application is highly interactive, user-friendly, and openly shared, facilitating interactions between spectroscopists and astronomers while advancing AIB analysis.

We will continue to enrich the database with our own data, but we also aim to collect available data from the literature with participation of spectroscopists producing either experimental or theoretical data. Some datasets are available in the literature concerning the evolution with temperature of the IR spectra of thermally excited neutral PAHs in the gas phase \citep{joblin1992, joblin1994, farouki2025}. Others were recorded only at a single temperature (NIST webbook), in particular, room temperature \citep[NIST WebBook and other works, e.g.][] {pirali2010, MartinDrumel2013} or at low temperature in jet-cooled experiments \citep{maltseva2016, lemmens2021}. For ionized species, measurements for energized species are coming from IRMPD spectroscopy \citep{oomens2006, zhen2017, wiersma2022} using the free electron laser FELIX. However, their interpretation to derive information on anharmonicity is complex \citep{oomens2006}. Laboratory data on UV-excited PAHs are very scarce, mostly limited to neutral species \citep{cook1996}, with only one study on pyrene cations \citep{kim2002}.  
Theoretical spectra as a function of temperature are also very limited and insufficiently tested \citep{basire2009, basire2010, joalland2010, simon2011, mackie2018b, chen2018ApJS, chen2018AandA,chen2019,tang2025,mai2025}. Therefore, both experimental and theoretical data are critically required to advance this study, which is central to the analysis of JWST observations and future IR facilities such as the Extremely Large Telescope.
 
Our work on pyrene and derivatives has allowed us to propose an identification for the carrier of the red component of the 3.4~$\mu$m band \citep{demyk2026}.
Addressing the analysis of AIBs in a comprehensive way requires a general understanding of the effect of temperature on IR spectra for different PAH sizes, structures, charges, and all bands. While we are not there yet, increasing the sample of species will help identify general trends, allowing us to move beyond the specificities of individual molecules. This would then strengthen the current effort to explore the chemical diversity of the AIB carriers using the NASA Ames PAH IR Spectroscopic Database and associated tools \citep[e.g.][]{bouwman2020,vats2023, ricca2026, maragkoudakis2026}.

\section*{Acknowledgments}
 This work was supported by the French space agency CNES under the project LAIBrary of the JWST-MIRI mission, the European Research Council under the European Union’s Seventh Framework Programme (ERC-2013-SyG, Grant agreement N$^{o}$610256 NANOCOSMOS), the Université of Toulouse and the Thematic Action “Physique et Chimie du Milieu Interstellaire” (PCMI) of the INSU Programme National “Astro”, with contributions from CNRS Physique, CNRS Chimie, CEA, and CNES.  The authors gratefully acknowledge Anthony Bonnamy and Loïc Noguès for their technical support on the ESPOIRS setup of the Nanograin platform. 




\section*{Data availability}
Both \texttt{cosmicPAHmfit} and \texttt{CosmicPAH\textendash IRDB} are accessible through the \texttt{Cosmic PAH portal}.\\
\texttt{cosmicPAHmfit} is an official PyPI package that can be downloaded using 
\texttt{pip install} \texttt{cosmicPAHmfit}. 
It was developed and thoroughly tested on Python version 3.12, it is likely (but not guaranteed) to work with newer Python versions. It is recommended to use a virtual environment. The user manual is available by downloading the application package or directly on the \href{https://cosmic-pah.irap.omp.eu/}{\texttt{Cosmic PAH portal}}. The help page of \texttt{CosmicPAH\textendash IRDB} is accessible directly from the database website.


\printcredits

\bibliographystyle{cas-model2-names}




\end{document}